\documentclass[aps,pra,twocolumn,10pt,citeautoscript,longbibliography]{revtex4-1}
\usepackage{amssymb}
\usepackage{amsmath}
\usepackage[pdftex]{graphicx}
\usepackage{dcolumn}
\usepackage{bm}
\usepackage{textcomp}
\usepackage{sidecap}
\usepackage{soul}

\usepackage[svgnames]{xcolor}
\definecolor{trueblue}{rgb}{0.0, 0.45, 0.81}
\definecolor{crimsonglory}{rgb}{0.75, 0.0, 0.2}

\usepackage{hyperref}
\hypersetup{colorlinks,linkcolor={trueblue},citecolor={crimsonglory},urlcolor={crimsonglory}}

\begin{document}

\title{Fingerprints of a position-dependent Fermi velocity on scanning tunnelling spectra of strained graphene}

\author{M. Oliva-Leyva$^1$}
\email{mauriceoliva.cu@gmail.com}
\author{J. E. Barrios-Vargas$^2$}
\email{j.e.barrios@gmail.com}
\author{Chumin Wang$^1$}
\email{chumin@unam.mx}

\affiliation{$^1$Instituto de Investigaciones en Materiales, Universidad Nacional Aut\'{o}noma de M\'{e}xico, Apartado Postal 70-360, 04510 Mexico City, Mexico.}

\affiliation{$^2$Departamento de F\'{i}sica, Facultad de Ciencias Fı\'{i}sicas y Matem\'aticas, Universidad de Chile, Santiago, Chile.}


\begin{abstract}

Nonuniform strain in graphene induces a position dependence of the Fermi velocity, as recently demonstrated by scanning tunnelling spectroscopy experiments. In this work, we study the effects of a position-dependent Fermi velocity on the local density of states (LDOS) of strained graphene, without and with the presence of a uniform magnetic field. The variation of LDOS obtained from tight-binding calculations is successfully explained by analytical expressions derived within the Dirac approach. These expressions also rectify a rough Fermi velocity substitution used in the literature that neglects the strain-induced anisotropy. The reported analytical results could be useful for understanding the nonuniform strain effects on scanning tunnelling spectra of graphene, as well as when it is exposed to an external magnetic field. 


\end{abstract}


\maketitle

\section*{Introduction}

Unlike most of the crystals, graphene can be reversibly stretched beyond $10\,\%$. This unusual elastic response has made it suitable to modify its electronic and optical properties via strains, idea known as strain engineering \cite{Amorim2016,Naumis2017}. For instance, when graphene is uniformly deformed, its low-energy electronic band structure around the Dirac points becomes elliptical cones. This fact can be visualized as an anisotropy of the Fermi velocity \cite{Oliva2017a}. As a consequence, the optical conductivity of graphene under uniform strain results anisotropic \cite{Pellegrino2010,Oliva2014}, which produces a modulation of the optical transmittance as a function of the incident light polarization \cite{Pereira2010,Oliva2015b}. This strain sensitivity of the optical response of graphene has been experimentally observed \cite{Ni2014} and, as proposed, it could be utilized towards the design of novel ultra-thin optical devices and strain sensors \cite{Bae2013}. Furthermore, it has been recently shown that the Faraday (Kerr) effect in graphene can be modified by means of deformations \cite{Oliva2017b}.

Nonuniform strains constitute even more useful tools to archive new behaviors of graphene. For example, the emergence of a pseudomagnetic field caused by a nonuniform strain can produce a pseudoquantum Hall effect in absence of external magnetic field \cite{Guinea2010a,Guinea2010b}. Nowadays, signatures of such gauge field in the electronic transport properties of graphene are actively investigated \cite{Stegmann2016,Mikkel16b,Carrillo2016,Georgi2016,Jones2017,Mikkel2017}.  Moreover, nonuniform strains graphene opens new opportunities to investigate others striking behaviors such as fractal spectrum \cite{Naumis2014}, metal-insulator transition \cite{Tang2015}, superconducting states \cite{Kauppila2016} and magnetic phase transitions \cite{Brey2016}. Within the Dirac approximation, in addition to the mentioned pseudomagnetic field, nonuniform strains give rise another recognized effect: a position-dependent Fermi velocity (PDFV) \cite{FJ2012}. However, signatures of PDFV in the graphene physics have been less addressed, even though they are always present for any nonuniform strain. 

Given that STS spectra provide the local density of states (LDOS), which depends on the Fermi velocity $v_{0}$ as $\rho_{0}(E)\sim \vert E\vert /v_{0}^{2}$ for pristine graphene, the slopes of V-shaped STS spectra present variations at different positions of the sample if the Fermi velocity is spatially varying. Based on this idea, evidence of the PDFV effect in strained graphene has been provided in a few experiments through scanning tunnelling spectroscopy (STS) \cite{Hui2013,Jang2014}. However, to obtain a local measurement of the Fermi velocity, typically $v_{0}$ is replaced by $v(x)$ in $\rho_{0}(E)$ leading to $\rho(E,x)\sim \vert E\vert /v^{2}(x)$, where $x$ is the measured position across the strain direction. According to this substitution, Fermi velocities at two different positions, $v(x_{1})$ and $v(x_{2})$, are related by the expression, $v(x_{1})/v(x_{2})=[\mathcal{S}(x_{2})/\mathcal{S}(x_{1})]^{1/2}$, where $\mathcal{S}(x)$ is the STS spectrum slope at the position $x$~\cite{Jang2014}. A purpose of this work is to clarify that the appropriate expression is given by $v(x_{1})/v(x_{2})=\mathcal{S}(x_{2})/\mathcal{S}(x_{1})$, at least when a space dependent Fermi velocity is due to a nonuniform uniaxial strain.

From the quantum field theory it has been pointed out that a PDFV (in curved graphene) becomes spatial modulations of LDOS \cite{FJ2007}; nevertheless, a better description of strain-induced PDFV has been arrived at from low-energy expansions of the standard tight-binding model \cite{FJ2012,FJ2013,Zubkov2014,Oliva2015a}. The achievement of these last studies consists of determining the Fermi velocity tensor as a function of the position-dependent strain tensor. This fact has allowed to approximately calculate within the Dirac model the PDFV effect on the LDOS and, therefore, on STS measurements. However, the analytical expressions for the LDOS of strained graphene, reported in Ref. \cite{FJ2013}, have not been compared with results obtained from tight-binding calculations. Such a comparison will be presented in this article. 

Moreover, STS experiments in the presence of a magnetic field can also be used to reveal local variations of the Fermi velocity, as performed in randomly strained graphene \cite{Andrei2011} as well as on the surface of a complex topological insulator \cite{Storz2016}. Here, we report the first detailed study, to our best knowledge, of the PDFV effect on Landau-level spectroscopy using both approaches, tight-binding model and Dirac approximation, in order to provide a better understanding and a more complete theoretical framework for these types of experiments carried out in strained graphene under an external magnetic field.

\begin{figure*}
\includegraphics[width=0.75\textwidth]{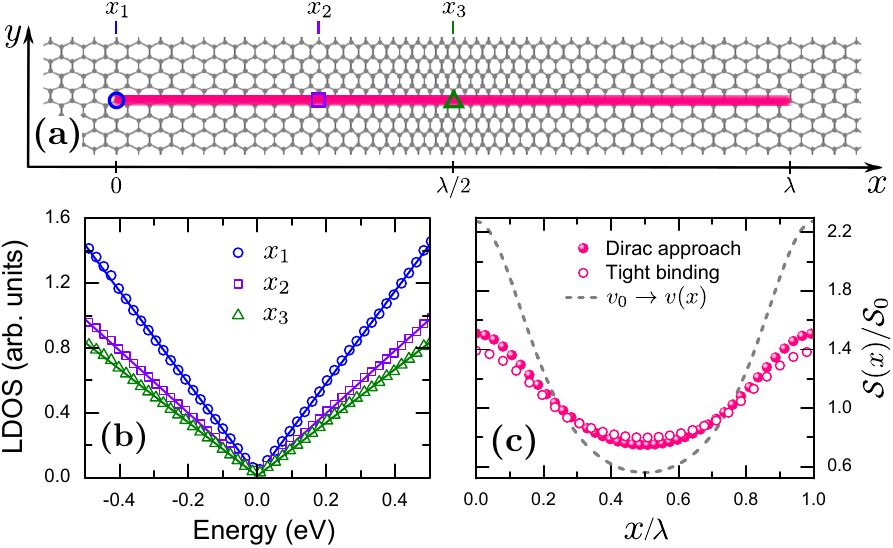}
\caption{\label{fig1}\textbf{(a)} Schematic representation of an oscillating displacement field with wavelength $\lambda$ along the zigzag direction of graphene. \textbf{(b)} Tight-binding results of the local density of states (LDOS) at three positions illustrated in panel (a). Solid lines obtained from fitting indicate the LDOS slope, $\mathcal{S}(x)$, at distinct sites. \textbf{(c)} Strain-induced variation of positive $\mathcal{S}(x)$ along the pink path in panel (a), according to three different approaches denoted in the figure. Results of panels (b) and (c) are obtained for $\lambda=660\sqrt{3}a_{0}$ and $2\pi u_{0}=0.1 \lambda$.}
\end{figure*}

\section{Position-dependent Fermi velocity effect on LDOS} 

For graphene, the electronic implications of strain can be investigated by means of the nearest-neighbor tight-binding Hamiltonian
\begin{equation}\label{TB}
\mathcal{H}=-\sum_{\langle i,j \rangle} t_{ij} c_{i}^{\dagger} c_{j} ,
\end{equation}
where the sum $\langle i,j \rangle$ runs over nearest neighbors and $c_{i}^{\dagger}$ $(c_{i})$ is the creation (annihilation) field operator at the
$i$th site. The strain-induced modification of the nearest-neighbor hopping parameter $t_{ij}$ is captured by \cite{Pereira2009}
\begin{equation}\label{t}
t_{ij}=t_{0}\exp[-\beta(d_{ij}/a_{0}-1)],
\end{equation}
where $t_{0}=2.7\,\textrm{eV}$, $\beta=3.37$, $a_{0}=0.142\,\textrm{nm}$  is the interatomic distance for unstrained graphene, and $d_{ij}$ is the modified distance between atomic sites $i$ and $j$. It is relevant for the present discussion to note that the Hamiltonian~(\ref{TB}) with constant $t_{ij}$ is not capable to describe purely geometric effects on LDOS induced by nonuniform strain \cite{FJ2012}. For example, if one assumes $\beta=0$, Eq.~(\ref{TB}) becomes $\mathcal{H}=-t_{0}\sum_{\langle i,j \rangle} c_{i}^{\dagger} c_{j}$ which has the same eigenenergies and eigenfunctions of pristine graphene even when the atoms move from their equilibrium positions.

In order to isolate PDFV effects, we consider a nonuniform uniaxial strain along the zigzag crystalline orientation, as shown in Fig.~\ref{fig1}(a), which is generated by a displacement field of the form $\bm{u}(x,y) = u_0 \sin(2\pi x/\lambda) \bm{e}_{x}$, with $\bm{e}_{x}$ being the unit vector in the $x$-direction. The parameters $\lambda$  and $u_{0}$ fulfill conditions $a_{0}/\lambda\ll 1$ and $2\pi u_{0}/\lambda\ll 1$, allowing the comparison between discrete and continuous models. In fact, this particular nonuniform strain does not produce pseudomagnetic field $B_{s}$. Because according to the considered displacement field $\bm{u}$, the components of the strain tensor defined by $\epsilon_{ij}=(\partial_{i}u_{j}+\partial_{j}u_{i})/2$ result $\epsilon_{xx}(x)=(2\pi u_0/\lambda)\cos(2\pi x/\lambda)$ and $\epsilon_{yy}=\epsilon_{xy}=0$; at the same time $B_{s}$ is related to the strain tensor through $B_{s} \sim (\partial_{y}\epsilon_{xx}-\partial_{y}\epsilon_{yy}+2\partial_{x}\epsilon_{xy})$ \cite{Guinea2010a}, therefore $B_{s}=0$. It is worth noting that for an out-of-plane displacement field $h(x)$, e.g. a ripple along the zigzag direction, the physical situation is essentially the same as that studied here, since one has only PDFV effects within the linear approximation \cite{Oliva2015a}. 

Within the tight-binding model, we calculate the LDOS at the atomic site $n$ through $\rho(E,n)=(-1/\pi)\text{Im}[G^{+}(n,n;E)]$, where $G^{+}(n,m;E)$ is the retarded Green's function evaluated using a uniform Monkhorst-Pack grid~\footnote{We set the Monkhorst-Pack grid-spacing equal to $0.0013\,\text{\AA}^{-1}$}. Given the considered displacement field, the LDOS depends only the $x$-coordinate and, besides, it results $\lambda$-periodic if $\lambda$ is a multiple of $\sqrt{3}a_{0}$. 

Figure~\ref{fig1}(b) shows the LDOS at low energies, for $\lambda=660\sqrt{3}a_{0}$ and $2\pi u_{0}=0.1 \lambda$, at three distinct sites of the strained graphene sample which are labelled as $x_{1}$, $x_{2}$ and $x_{3}$, respectively. At $x_{1}$ the sample has a maximum local stretching along the zigzag direction of $10\,\%$ ($\epsilon_{xx}=0.1$), whereas that at $x_{3}$ it has a maximum local shrinking of $10\,\%$ ($\epsilon_{xx}=-0.1$). In contrast, at $x_{2}$ the local strain is close to zero ($\epsilon_{xx}\approx 0$). This change of strain along of the zigzag direction induces a variation of the LDOS observed in Fig.~\ref{fig1}(b). In brief, the V-shape of LDOS at low energies is more widely-opened in the shrinked region than in the stretched one. 

This behavior is quantitatively registered by Fig.\ref{fig1}(c), where we present the positive slope of the V-shaped LDOS $\mathcal{S}(x)$, in units of the slope $\mathcal{S}_0$ for unstrained graphene, as a function of the position $x$ from three different approaches. The open circles correspond to our tight-binding calculations of $\mathcal{S}(x)$, whereas the solid circles are given by 
$\mathcal{S}(x)=\mathcal{S}_{0}/[1-\beta \epsilon_{xx}(x)]$, according to the Dirac approach \cite{FJ2013}. Good agreement is observed for these two approaches, confirming that the LDOS variation is indeed induced by PDFV. 

Let us explain the essence of the analytical expression derived within the Dirac approximation. For graphene under uniform strain, i.e. non-position dependent strain, the LDOS results $\rho(E)=\rho_{0}(E)/\text{det}(v_{ij}/v_{0})$, where $v_{ij}=v_{0}(1-\beta\epsilon_{ij}+\epsilon_{ij})$ is the strain-induced Fermi velocity tensor \cite{Oliva2014} whose $\beta$-independent term $v_{0}\epsilon_{ij}$ is purely a geometric consequence due to the lattice deformation \cite{Oliva2015a}. Then for a sufficiently smooth spatially-varying strain, the LDOS can be approximately calculated  by making the substitution $\epsilon_{ij}\rightarrow \epsilon_{ij}(x,y)$. In consequence, for our problem $\rho(E,x)=\rho_{0}(E)/[1-\beta \epsilon_{xx}(x)+\epsilon_{xx}(x)]$, but disregarding geometric effects (given by the $\beta$-independent term) in order to compare with the tight-binding results, one finally get $\mathcal{S}(x)=\mathcal{S}_{0}/[1-\beta \epsilon_{xx}(x)]$.

In other words, the considered strain only modifies the Fermi velocity in the $x$-direction, whereas the Fermi velocity in the $y$-direction remains equal to $v_{0}$. Hence, the substitution $v_{0}\rightarrow v(x)$ in $\rho_{0}(E)=2\vert E\vert /(\pi\hbar^{2}v_{0}^{2})$ is not appropriate to obtain $\rho_{0}(E,x)$ because when making such an substitution, one would be wrongly assuming that both components of the velocity are equally modified by strain. To visualize this fact, in Fig.\ref{fig1}(c) we illustrate with dash line the variation of the slope derived by replacing $v_{0}$ in $\rho_{0}(E)$ by $v_{0}[1-\beta\epsilon_{xx}(x)]$, which remarkably differs from the ones obtained from the tight-binding and Dirac approaches. Therefore, to obtain a more accurate LDOS of graphene under a nonuniform uniaxial strain (e.g. a ripple as considered in Ref.~\cite{Jang2014}) the appropriate replacement should be $v_{0}^{2}\rightarrow v_{0}v(x)$ in $\rho_{0}(E)$, where $v(x)$ is the Fermi velocity along the strain direction. In consequence, one gets $\rho(E,x)=\rho_{0}(E)v_{0}/v(x)$, whence the LDOS slopes at two different positions, $\mathcal{S}(x_{1})$ and $\mathcal{S}(x_{2})$, are related by the expression
\begin{equation}\label{S}
\mathcal{S}(x_{1})/\mathcal{S}(x_{2})=v(x_{2})/v(x_{1}).
\end{equation}

\section{Position-dependent Fermi velocity effect on LDOS in the presence of magnetic field} 

We now add a uniform magnetic field $B\bm{e}_{z}$ to our previously discussed problem of strained graphene.  

Within the tight-binding model, the Hamiltonian has the form (\ref{TB}), but now the hopping parameter $t_{ij}$ is evaluated by a generalized expression of Eq.~(\ref{t}) as
\begin{equation}\label{tB}
t_{ij}=t_{0}\exp[-\beta(d_{ij}/a_{0}-1)]\exp[i\phi_{ij}],
\end{equation}
in which the magnetic field effect is introduced by the Peierls phase $\phi_{ij}$, according to \cite{Lutinger51,BookFoa}  
\begin{equation}
\phi_{ij}=\frac{2\pi}{\phi_{0}}\int_{\bm{r}_{i}}^{\bm{r}_{j}}\bm{A}(\bm{r})\cdot d\bm{r},
\end{equation}
where $\phi_{0}$ is the magnetic flux quantum, $\bm{A}(\bm{r})$ is the vector potential and $\bm{r}_{i}$ ($\bm{r}_{j}$) denotes the modified position of site $i$ ($j$). Unlike the case without magnetic field, the tight-binding Hamiltonian in the presence of magnetic field captures purely geometric effects due to strain. Even for $\beta=0$, the resulting hopping parameter $t_{ij}=t_{0}\exp[i\phi_{ij}]$ depends on the position through the Peierls phase, leading to a spatial modulation of the LDOS. 

\begin{figure*}
\includegraphics[width=0.75\textwidth]{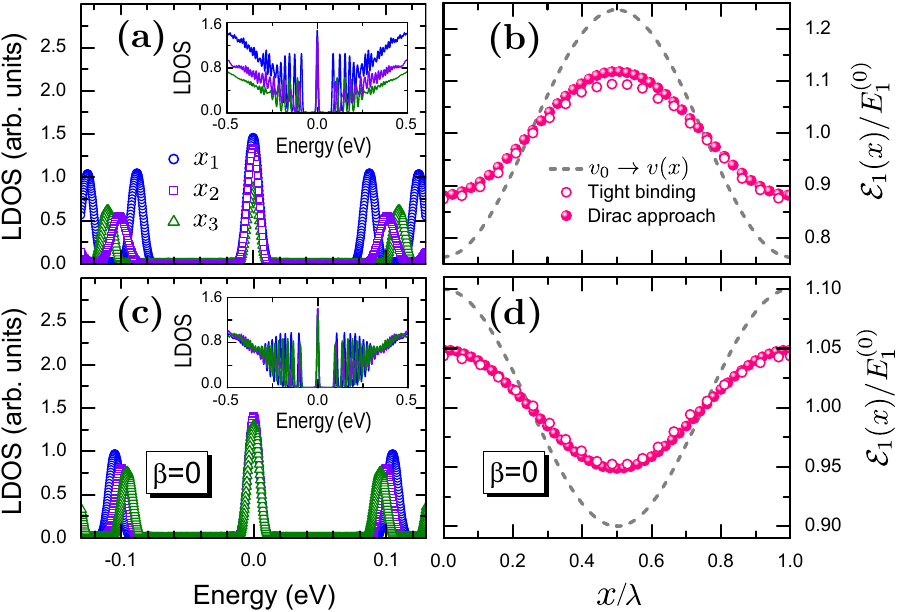}
\caption{\label{fig2}  Tight-binding calculations of the local density of states (LDOS) for graphene under a nonuniform uniaxial strain, as illustrated in Fig.~\ref{fig1}(a), and in the presence of a magnetic field $B$. \textbf{(a)} LDOS at three positions as indicated in Fig.~\ref{fig1}(a) with the corresponding colors. 
\textbf{(b)} First peak of the LDOS, $\mathcal{E}_{1}(x)$, as function of the position $x$ along the strain direction, according to three different approaches. Panels (c) and (d) are analogous to panels (a) and (b), respectively, but assuming $\beta=0$. Insets: LDOS over an extended energy range. Parameters: $\lambda=660\sqrt{3}a_{0}$, $2\pi u_{0}=0.1 \lambda$ and $B=10\,\textrm{T}$.}
\end{figure*}

Figure~\ref{fig2}(a) is analogous to Fig.~\ref{fig1}(b), but the shown LDOS were calculated by assuming a magnetic field of magnitude $B=10\,\text{T}$. The most remarkable feature of Fig.~\ref{fig2}(a) is the presence of a series of well defined peaks. For pristine graphene, the LDOS presents such peaks at the Landau level energies, given by $E^{(0)}_{n}=\pm\sqrt{2 e\hbar v_{0}^{2}B n}$ \cite{Andrei2012}. For example, the first (positive) peak is at $E^{(0)}_{1}\approx 0.1\,\text{eV}$ for $B=10\,\text{T}$. In Fig.~\ref{fig2}(a) for nonuniformly strained graphene, the first peak of the LDOS, located at energy $\mathcal{E}_{1}(x)$, is around $0.1\,\text{eV}$ but it depends on the position $x$. For instance, $\mathcal{E}_{1}(x_1) < 0.1\,\text{eV}$ in the stretched region, $\mathcal{E}_{1}(x_3) > 0.1\,\text{eV}$ in the shrinked region, and $\mathcal{E}_{1}(x_2)\approx 0.1\,\text{eV}$, where local strain is approximately zero. This variation of $\mathcal{E}_{1}(x)$ with the position $x$ is quantitatively displayed by open pink circles in Fig.~\ref{fig2}(b), according to our tight-binding calculations. Moreover, to visualize the purely geometric effects due to strain, in Fig.~\ref{fig2}(c) we show the LDOS, as analogously made in Fig.~\ref{fig2}(a), but obtained by assuming $\beta=0$ in Eq.~(\ref{tB}). For this hypothetical case, the dependence of $\mathcal{E}_{1}(x)$ as function of $x$ is opposite to that of the realistic case with $\beta=3.37$, which can be clearly noted by comparing Fig.~\ref{fig2}(b) and Fig.~\ref{fig2}(d).

Let us provide an explanation to the tight-binding results of $\mathcal{E}_{1}(x)$ from the Dirac approximation in terms of PDFV effects on the LDOS. For graphene under uniform strain, the Landau level energies are given by $E_{n}=E_{n}^{(0)}\sqrt{\text{det}(v_{ij}/v_{0})}$, with $v_{ij}=v_{0}(1-\beta\epsilon_{ij}+\epsilon_{ij})$ \cite{Oliva2017b}. Hence, the first peak of the LDOS for uniformly strained graphene should be at $E_{1}=E_{1}^{(0)}[1-(\beta-1)\text{tr}(\epsilon_{ij})/2]$, up to first-order in the strain tensor. Then, for a sufficiently smooth spatially-varying strain, $\mathcal{E}_{1}(x,y)$ can be approximately estimated by making the substitution $\epsilon_{ij}\rightarrow \epsilon_{ij}(x,y)$ in $E_{1}$, which leads to
\begin{equation}\label{E1}
\mathcal{E}_{1}(x,y)=E_{1}^{(0)}\left\lbrace 1-\frac{\beta-1}{2}\mbox{tr}[\epsilon_{ij}(x,y)]\right\rbrace.
\end{equation}
Thus, at a locally dilated region ($\text{tr}[\epsilon_{ij}(x,y)]>0$) the  first positive peak of the LDOS is shifted to the left of $\vert E_{1}^{(0)}\vert$, whereas at a locally compressed region ($\text{tr}[\epsilon_{ij}(x,y)]<0$) the first positive peak of the LDOS is shifted to the right. 

From Eq.~(\ref{E1}), for a nonuniform strain as illustrated in Fig.~\ref{fig1}(a), it follows that $\mathcal{E}_{1}(x)=E_{1}^{(0)}[1-(\beta-1) \epsilon_{xx}(x)/2]$. In Fig.~\ref{fig2}(b) with $\beta=3.37$ and Fig.~\ref{fig2}(d) with $\beta=0$, it can be observed a good agreement between the results predicted by the last analytical expression, according to the Dirac approach, and those obtained from the tight-binding model. This fact confirms the concept of a PDFV for the understanding and description of LDOS variations induced by a nonuniform uniaxial strain. Moreover, in Figs.~\ref{fig2}(b,d) we illustrate by the dashed lines the consequence of  substitution $v_{0}\rightarrow v(x)$ in $E_{1}^{(0)}$, which leads to results notably different from those obtained by the tight-binding and Dirac approaches. In short, to evaluate approximately $\mathcal{E}_{1}(x)$ for graphene under a nonuniform uniaxial strain (e.g. a ripple) and in the presence of a uniform magnetic field, the appropriate replacement should be $v_{0}^{2}\rightarrow v_{0}v(x)$ in $E_{1}^{(0)}$, hence one gets $\mathcal{E}_{1}(x)=E_{1}^{(0)}\sqrt{v(x)/v_{0}}$, keeping in mind that $v(x)$ is the Fermi velocity along the strain direction. Therefore, using Landau-level spectroscopy measurements, Fermi velocities at two different positions, $v(x_{1})$ and $v(x_{2})$, are related by 
\begin{equation}\label{E}
v(x_{1})/v(x_{2})=\left[\mathcal{E}_{1}(x_{1})/\mathcal{E}_{1}(x_{2})\right]^2.
\end{equation}

Equation~(\ref{E}) is certainly limited to a nonuniform uniaxial strain along the zigzag direction. For a situation beyond uniaxial strain, the PDFV effect on the LDOS peaks can be quantified by a more general expression as Eq.~(\ref{E1}), which is valid whenever the strain-induced pseudomagnetic field $B_{s}(x,y)$ fulfills the condition $B_{s}(x,y)/B\ll\mbox{tr}[\epsilon_{ij}(x,y)]$. 

On the other hand, as illustrated in the inset of Fig.~\ref{fig2}(a), the LDOS presents an inclination (slope) over an extended energy range. Furthermore, this inclination depends on the position in the same manner as occurred in Fig.~\ref{fig1}(b), which can be explained as a PDFV effect. For the case $\beta=0$, see inset of Fig.~\ref{fig2}(c), the mentioned position dependence of the LDOS inclination is less remarked because the PDFV effect is only introduced through the Peierls phase.

\section*{Conclusions} 

In closing, we have presented a numerical and analytical study of the LDOS of graphene under nonuniform uniaxial strain, either in absence or in presence of a uniform magnetic field. Our tight-binding results of the LDOS have been successfully explained by analytical expressions derived from the Dirac approximation in term of a PDVF. Moreover, we have clarified that the replacement $v_{0}\rightarrow v(x)$ in expressions of pristine graphene (e.g. $\rho_{0}(E)\sim \vert E\vert /v_{0}^{2}$) could be not appropriate to evaluate PDFV effects on LDOS since such rough approach disregards the strain-induced anisotropy of the Fermi velocity. In consequence, the analytical expressions (\ref{S}) and (\ref{E}) for the LDOS can be useful to describe appropriately effects of nonuniform uniaxial strains (e.g. ripples along the zigzag direction) on STS experiments of graphene, without and with the presence of a uniform magnetic field. It is important to mention that the analytical results reported in this article are valid for $\lambda\gg\ell_{B}\gg a_{0}$, where $\ell_{B}=\sqrt{\hbar/(eB)}$ is the magnetic length. In addition to the contribution of this work for understanding PDFV effects on Landau-level spectroscopy measurements of strained graphene, our results also suggest that PDFV effects should be considered in a complete description of transport signatures of strain-induced pseudomagnetic fields. 
 
\begin{acknowledgments}
This work has been partially supported by CONACyT of Mexico through Project 252943, and by PAPIIT of Universidad Nacional Aut\'onoma de M\'exico (UNAM) through Project IN106317. Computations were performed at Miztli of UNAM. M.O.L. acknowledges the postdoctoral fellowship from DGAPA-UNAM. J.E.B.V. acknowledges support from FondeCyT-Postdoctoral 3170126.
\end{acknowledgments}

\bibliography{biblioStrainedGraphene}

\end{document}